\documentclass[twocolumn]{aastex631}

\usepackage[english]{babel}
\usepackage{amsmath}
\usepackage{amssymb}
\usepackage{graphicx}
\usepackage{subfigure}
\usepackage[normalem]{ulem}
\usepackage{enumerate}
\usepackage{booktabs}

\usepackage{verbatim}

\usepackage{color}
\usepackage{multirow}
\usepackage{mathtools}
\usepackage{epstopdf}
\usepackage{tabularx}

\usepackage{url}
\usepackage{xcolor}

\def\apjl{ApJL}
\def\apjs{ApJS}
\def\aap{A\&A}

\def\be{\begin{equation}} 
\def\ee{\end{equation}} 
\def\ba{\begin{eqnarray}} 
\def\ea{\end{eqnarray}}

\def\kms{\,{\rm {km\, s^{-1}}}} 
 
\def\msun{{\Msun}}

\def\gsim{\lower.5ex\hbox{\gtsima}} 
\def\lsim{\lower.5ex\hbox{\ltsima}} \def\gtsima{$\; \buildrel > \over 
\sim \;$} \def\ltsima{$\; \buildrel < \over \sim \;$} \def\prosima{$\; 
\buildrel \propto \over \sim \;$} \def\gsim{\lower.5ex\hbox{\gtsima}} 
\def\lsim{\lower.5ex\hbox{\ltsima}} 
\def\simgt{\lower.5ex\hbox{\gtsima}} 
\def\simlt{\lower.5ex\hbox{\ltsima}} 
\def\simpr{\lower.5ex\hbox{\prosima}}   
  
 \def\gtsima{$\; \buildrel > \over \sim \;$} 
\def\ltsima{$\; \buildrel < \over \sim \;$} 
\def\gsim{\lower.5ex\hbox{\gtsima}} 
\def\lsim{\lower.5ex\hbox{\ltsima}} 
\def\simgt{\lower.5ex\hbox{\gtsima}} 
\def\simlt{\lower.5ex\hbox{\ltsima}} 
\def\simpr{\lower.5ex\hbox{\prosima}}

\def\msun{\,{\rm \Msun}}

\def\E3{{\cal E}_{\rm g}^{III}}

\def\r12{r_{1/2}} 
\def\x12{x_{1/2}} 
\def\v12{v_{1/2}}

\newcommand\code[1]{\textsc{\MakeLowercase{#1}}}

\def\nh2{n_{\rm H2}}
\def\fh2{f_{\rm H2}}

\def\angstrom{\textrm{A\kern -1.3ex\raisebox{0.6ex}{$^\circ$}}}

\def\msun{{\rm M}_{\odot}}
\def\zsun{{\rm Z}_{\odot}}

\def\msunyr{\msun\,{\rm yr}^{-1}}

\makeatletter
\def\@hex@@Hex#1%
 {\if a#1A\else \if b#1B\else \if c#1C\else \if d#1D\else
  \if e#1E\else \if f#1F\else #1\fi\fi\fi\fi\fi\fi \@hex@Hex}
\makeatother

\definecolor{apcolor}{HTML}{b3003b}
\definecolor{cbcolor}{HTML}{ff0f00}
\definecolor{afcolor}{HTML}{b3443c}
\definecolor{vgcolor}{HTML}{8F00FF}
\definecolor{tbdcolor}{HTML}{E8A95E}
\definecolor{stefcolor}{HTML}{0047ab}

\shorttitle{SN quenching at high-$z$}
\shortauthors{Gelli et al.}

\begin{document}

\title{Can supernovae quench star formation in high-$z$ galaxies?}

\correspondingauthor{Viola Gelli}
\email{viola.gelli@unifi.it}

\author[0000-0001-5487-0392]{Viola Gelli}
\author[0000-0001-7298-2478]{Stefania Salvadori}
\affiliation{Dipartimento di Fisica e Astronomia, Universit\'{a} degli Studi di Firenze, via G. Sansone 1, 50019, Sesto Fiorentino, Italy}
\affiliation{INAF/Osservatorio Astrofisico di Arcetri, Largo E. Fermi 5, I-50125, Firenze, Italy}
\author[0000-0002-9400-7312]{Andrea Ferrara}
\author[0000-0002-7129-5761]{Andrea Pallottini}
\affiliation{Scuola Normale Superiore, Piazza dei Cavalieri 7, I-56126 Pisa, Italy}

\begin{abstract}
JWST is providing the unique opportunity to directly study feedback processes regulating star formation (SF) in early galaxies. The two $z>5$ quiescent systems (JADES-GS-z7-01-QU and MACS0417-z5BBG) detected so far show a recent starburst after which SF is suppressed.
To clarify whether such quenching is due to supernova (SN) feedback, we have developed a minimal physical model. We derive a condition on the minimum star formation rate, $\rm SFR_{min}$, lasting for a time interval $\Delta t_{b}$, required to quench SF in a galaxy at redshift $z$, with gas metallicity $Z$, and hosted by a halo of mass $M_h$. We find that lower $(z, Z, M_h)$ systems are more easily quenched.
We then apply the condition to JADES-GS-z7-01-QU ($z=7.3$, $M_\star=10^{8.6} M_\odot$) and MACS0417-z5BBG ($z=5.2$, $M_\star=10^{7.6} M_\odot$), and find that SN feedback largely  \textit{fails} to reproduce the observed quenched SF history. Alternatively, we suggest that SF is rapidly suppressed by radiation-driven dusty outflows sustained by the high specific SFR (43 and 25 Gyr$^{-1}$, respectively) of the two galaxies. 
Our model provides a simple tool to interpret the SF histories of post-starburst galaxies, and unravel quenching mechanisms from incoming JWST data.
\end{abstract}

\keywords{High-redshift galaxies, Galaxy evolution, Galaxy quenching, Cosmology}

\section{Introduction}

Galaxy formation, growth and evolution depend on the complex interplay of various physical mechanisms. Gas cooling leading to star formation (SF) can be counteracted by a variety of feedback processes that could push galaxies into temporary or more permanent states of \textit{quiescence}, i.e. extremely low or suppressed SF activity, especially at cosmic dawn when their SF occurs in a particularly bursty fashion \citep[e.g.][]{Pallottini23, Sun23}. Exploring and understanding the physical mechanisms behind SF \textit{quenching} is of fundamental importance to understand how galaxies evolve through cosmic times. 

A wide diversity of both internal and external physical processes, acting on different timescales and mass ranges, can be invoked to explain galaxy quenching.
Environmental effects such as ram pressure stripping and tidal interactions are typically associated with quenching occurring over long timescales ($>100$~Myr, e.g. \citealt{Emerick2016, Williams21_almaquiesc, Boselli2022}). 
\begin{figure*}[t!]
\centering
\includegraphics[width=\textwidth]{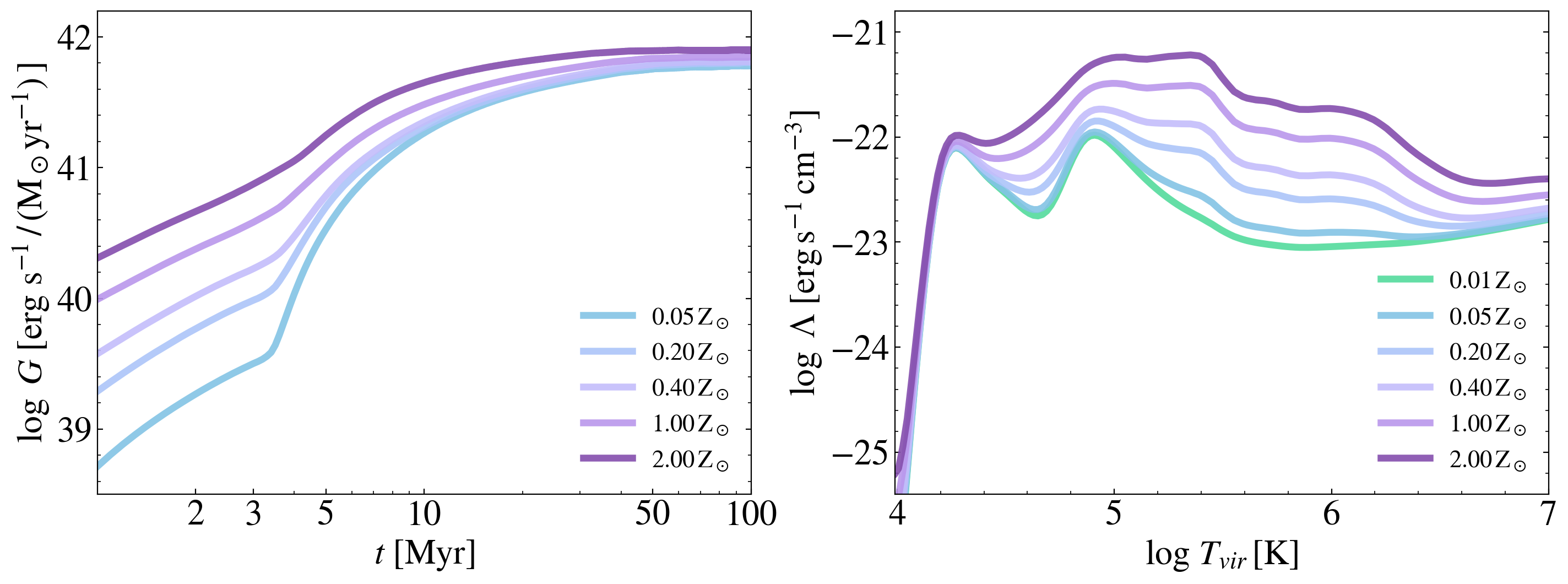}
\caption{
\textit{Left:} Energy input rate ($G$, Eq. \ref{eq:E+}) as a function of time for a continuous burst of star formation with $\rm SFR=1\, \msunyr$ derived with \code{starburst99} \citep{starburst99}.
\textit{Right:} cooling functions ($\Lambda$) in the interstellar medium as a function of temperature ($T_{\rm vir}$) derived with \code{KROME} \citep{krome}.
In both panels the different curves identify the different metallicity of stars and gas, respectively.
\label{fig:coolingrate}
}
\end{figure*}
Conversely, a more rapid quenching ($<50$~Myr) requires the action of internal physical mechanisms. Among these, feedback from active galactic nuclei (AGN) is typically invoked to explain the quenching of massive galaxies ($M_\star > 10^{10.5}\msun$) populating the high-mass end of the observed galaxy mass function. Stellar feedback driven by supernova (SN) explosions, is instead crucial to quench galaxies in the low-mass regime \citep[e.g.][]{FerraraTolstoy2000,Salvadori15}.
Indeed, these energetic explosions resulting from the death of massive stars may drastically influence low-mass systems with shallow potential wells, making it challenging to retain the newly metal-enriched gas within the galaxy itself \citep[e.g.][]{Gelli20}.

The vast majority of the quiescent galaxies that have been observed up to date are massive \citep[e.g.][]{Carnall20, Valentino20, Santini21}, but the \textit{James Webb Space Telescope} (JWST) is now opening a new window on the first generations of faint low-mass galaxies populating the high-redshift Universe.
Among these, the discovery of the first $z>5$ low-mass quiescent galaxies have been reported: the $z=7.3$, $M_\star= 10^{8.6}\msun$ JADES-GS-z7-01-QU \citep{Looser23} and the $z=5.2$, $M_\star=10^{7.6}\msun$ MACS0417-z5BBG \citep{Strait23}.
Interestingly, both galaxies appear to be experiencing a post-starburst phase in which star formation has been rapidly suppressed, implying that the quenching should be driven by internal feedback processes.

However in \cite{Gelli23} we showed that: (i) in order to reproduce the observed spectra of JADES-GS-z7-01-QU an \textit{abrupt quenching} of the simulated low-mass galaxy is required; and (ii) SNe alone \textit{cannot} induce such an abrupt SF quenching in the low-mass galaxies identified in our cosmological simulations \citep{Pallottini2022}. The same results have been later confirmed by \cite{Dome23} using the IllustrisTNG and VELA simulations.

Given the bursty nature of these low-mass high-$z$ galaxies and the intimate link between burst of SF and SN explosions, it is now of fundamental importance to understand, within a general physical framework, what is the impact of SNe on the evolution of galaxies with different masses and across different cosmic times.

The critical questions we want to address here are: (a) what is the role of SNe in driving the quenching of galaxies at different cosmic epochs? (b) Can they lead to a total quenching of star formation and under what conditions? To answer these questions in Sec.~\ref{sec:sn_quench} we derive a physical condition that can be used to infer the star formation rate (SFR) and burst duration $\Delta t_{b}$ required to quench SF as a function of galaxy properties. In Sec.~\ref{sec:results} we present the results and implications of the model. Sec.~\ref{sec:discussion} and Sec.~\ref{sec:conclusions} provide the discussion and conclusions, respectively.

\section{Supernova quenching}\label{sec:sn_quench}

We adopt a minimal approach\footnote{Throughout the paper, we assume a flat Universe with the following cosmological parameters: $\Omega_{m} = 0.3075$, $\Omega_{\Lambda} = 1- \Omega_{m}$, and $\Omega_{b} = 0.0486$,  $h=0.6774$, $\sigma_8=0.826$, where $\Omega_{m}$, $\Omega_{\Lambda}$, and $\Omega_{b}$ are the total matter, vacuum, and baryon densities, in units of the critical density; $h$ is the Hubble constant in units of $100\,\kms$, and $\sigma_8$ is the late-time fluctuation amplitude parameter \citep{planck:2015}.} to examine the impact of stellar winds and SN-feedback in leading to the quenching of high-$z$ galaxies after a burst of star formation.
Given that the currently observed high-$z$ quiescent galaxies show a very abrupt quenching, we model the star formation history (SFH) of a galaxy assuming a simple top-hat evolution: stars form in a burst of constant $\rm SFR$ over a time $\Delta t_{b}$, after which the SF drops to zero. 
Note that such kind of top-hat SFH is often assumed in Spectral Energy Distribution (SED) fitting models \citep[e.g. \code{bagpipes}][]{carnall_bagpipes} successfully reproducing the observed galaxies' properties. This top-hat approximation allows us to obtain a good estimate of the typical timescales required for a burst to produce the total quenching, although in a more realistic scenario the action of SN-feedback will cause a gradual decline of the SFR previous to the quenching \citep[e.g. see][]{Gelli23}. With this \textit{Ansatz}, the final stellar mass is then given by $M_\star = {\rm SFR}\times \Delta t_{b}$.

To explore the possibility of the galaxy of undergoing quenching induced by SNe, our model centers on a straightforward energetic argument: we assume that the star formation stops when the energy rate injected by SNe ($\dot{E}^{+}$) exceeds the cooling rate of the halo ($\dot{E}^{-}$). Under these conditions the halo gas remains hot/rarefied and is prevented to form stars. The condition for a SN-induced quenching is then:
\begin{equation}
    \dot{E}^{+} \geq \dot{E}^{-}\,.
    \label{eq:energyratecondition}
\end{equation}
In the following, we describe the assumptions adopted for SN energy input and gas cooling, detailing both terms of Eq. \ref{eq:energyratecondition}.

\subsection{SN energy input}

To compute the evolution of the stellar population in the galaxy, we use \code{starburst99} \citep{starburst99}, adopting the \code{Geneva} stellar tracks and assuming a Salpeter initial mass function (IMF) in the range $(1-100)~\msun$. Assuming a burst of constant $\rm SFR$, the energy released by stellar winds and SNe can be expressed as a function of time ($t$) and stellar metallicity ($Z_\star$). The energy rate evolution can be written as:
\begin{equation}
    \dot{E}^{+} = G(t, Z_\star) \, \rm SFR \,,
    \label{eq:E+}
\end{equation}
where the SFR is in units of $M_\odot \rm yr^{-1}$ and $G(t, Z_\star)$ is the energy input rate in [erg s$^{-1}/(M_\odot \rm yr^{-1})$] for a continuous burst of star formation with $\rm SFR=1~\msunyr$.
The energy rate as a function of time for a $\rm SFR=1~\msunyr$ burst is shown on the left in Fig.~\ref{fig:coolingrate} for different stellar metallicities. We see that the energy input rates increase for increasing metallicity. In particular, during the first $\leq 5$~Myr of evolution the rates of energy injected by metal-rich stellar populations ($Z_\star \geq \zsun$) exceed by more than an order of magnitude those of metal-poor ones, $Z_\star \leq 0.1\zsun$, while at later times their differences are less pronounced.
\begin{figure*}[t!]
\centering
\includegraphics[width=\textwidth]{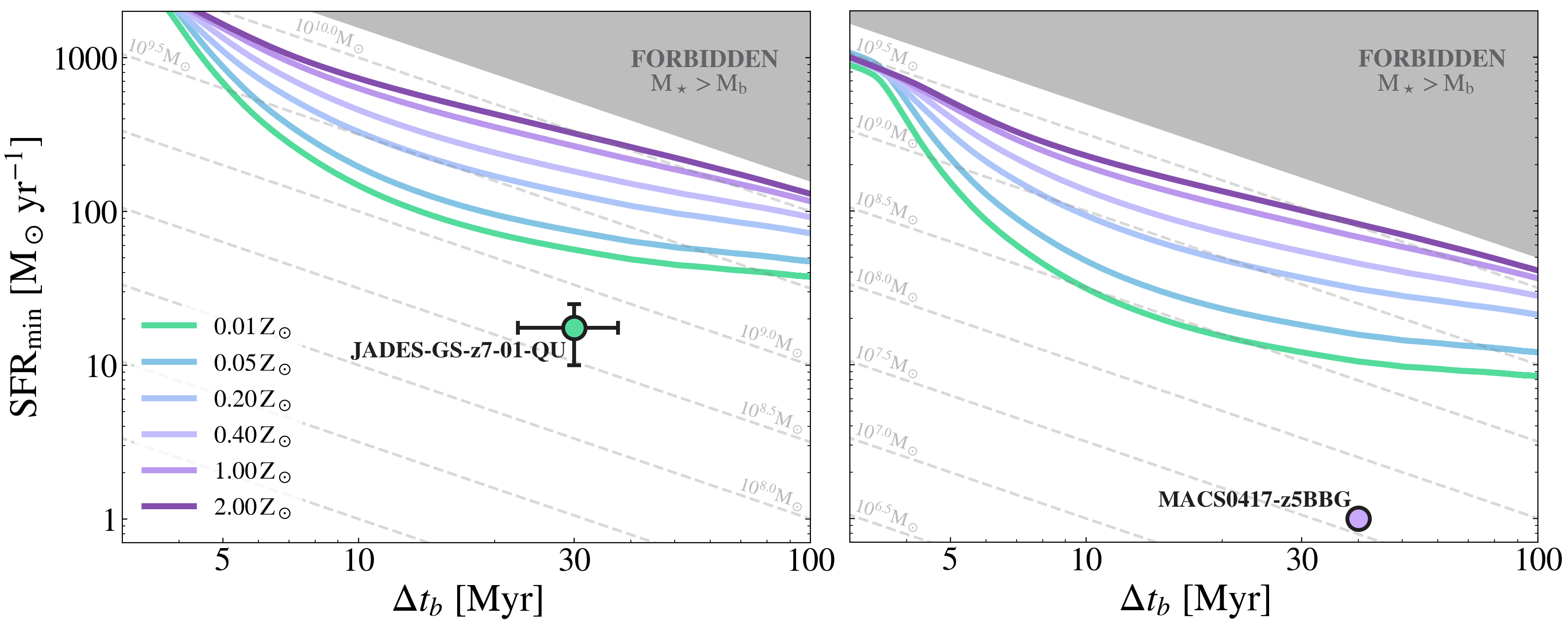}
\caption{
Minimum $\rm SFR$ necessary for SNe to quench SF following a burst of duration $\Delta t_{b}$, for different gas metallicities as indicated. Dashed grey lines indicate curves at constant stellar mass. We apply the model to two recently discovered quiescent galaxies.
\textit{Left panel}: predictions for JADES-GS-z7-01-QU \citep{Looser23} at $z=7.3$; the halo mass (derived from \citealt{Behroozi19}) is $M_h=10^{11}\msun$ corresponding to $T_{vir} = 10^{5.9}K$.
\textit{Right}: Same for MACS0417-z5BBG \citep{Strait23} at $z=5.2$, $M_h=10^{10.5}\msun$ and $T_{vir} = 10^{5.5}$ K.
The data points for each galaxy are obtained from SED fitting by approximating their SFHs prior to the quenching as a top-hat. Their color corresponds to the observed metallicity, i.e. $Z_\star=0.01\zsun$ for JADES-GS-z7-01-QU, and  $Z_\star=\zsun$ for MACS0417-z5BBG.
\label{fig:sfr_dt_jades}
}
\end{figure*}

\subsection{Gas cooling rate}

The most conservative assumption that can be made to impose the quenching of star formation in a galaxy consists in requiring that energy input from SNe heats all the halo gas above the virial temperature, $T_{vir}$, effectively counteracting cooling losses. This translates in expressing the cooling rate in Eq.~\ref{eq:energyratecondition} as:
\begin{equation}
    \dot{E}^{-} = \frac{M_g}{\mu m_p} k_B T_{vir} \frac{1}{t_{cool}},
    \label{eq:e-basic}
\end{equation}
where $M_g$ is the total gas mass of the galaxy, $t_{cool}$ is the gas cooling time, $\mu = 0.6$ is the mean molecular weight for ionized gas.

Considering that the baryonic mass is $M_b  = (\Omega_b/\Omega_m) M_h \equiv f_b M_h$,  and further expressing the gas mass as $M_g = M_{b} - M_\star$, Eq.~\ref{eq:e-basic} can be rewritten as:
\begin{equation}
    \dot{E}^{-} =  \frac{3}{2} \frac{n \Lambda(T_{vir}, Z)}{\mu m_p} [f_b M_h(z, T_{vir}) - {\rm SFR}\,\Delta t_{b}]\,, \label{eq:E-}
\end{equation}
where $\Lambda(T_{vir}, Z)$ is the cooling rate, $n$ (and $Z$) the gas number density (and metallicity). The gas density is  $\rho= \mu m_p n = 18\pi^{2}\Omega_b \rho_c(z)$, where $\rho_c(z)$ is the critical density at redshift $z$.
The cooling function is computed using \code{KROME} \citep{krome}, and it is shown on the right panel of Fig.~\ref{fig:coolingrate} as a function of $T_{vir}$ for different metallicities. For virial temperatures $T_{vir} \geq 10^{4.2}$~K we see that the higher the gas metallicity the higher the gas cooling rate. The differences are maximal at $T_{vir} \approx 10^{5.5}$~K, where the cooling rate of metal-rich gas exceed by almost two orders of magnitude the one of metal-poor gas, $Z \leq 0.01\zsun$.

\begin{table}[t!]
\centering
    \caption{
    For the two observed quiescent galaxies JADES-GS-z7-01-QU and MACS0417-z5BBG the table shows the following quantities: redshift $z$, stellar mass $M_\star$, metallicity $Z_\star$, halo mass $M_h$ (derived from \citealt{Behroozi19}),  SFR, burst duration $\Delta t_b$, specific star formation rate at the end of the burst sSFR, and dust extinction, $A_V$.
    \label{tab:gal_properties}
    }
    \begin{tabular}{ lcc } 
        \tableline 
        Parameter& JADES-GS-z7-01-QU & MACS0417-z5BBG \\ [0.6ex] 
        \tableline
        $z$                     &  7.3  &  5.2 \\ [0.6ex] 
        $\log M_\star /\msun$   &  8.6  &  7.6 \\ [0.6ex] 
        $Z_\star /\zsun$        &  0.01 &  1.0 \\ [0.6ex] 
        $\log M_h /\msun$       & 11.0  & 10.5 \\ [0.6ex] 
        $\rm SFR\, /\msunyr$    & 17.0  &  1.0 \\ [0.6ex] 
        $\Delta t_b\, /\rm Myr$ & 30.0  & 40.0 \\ [0.6ex] 
        $\rm sSFR\, /Gyr^{-1}$  & 43.0  & 25.0 \\ [0.6ex] 
        $A_V$                   &   0.1 & 0.19 \\ [0.6ex] 
        \tableline
    \end{tabular}
\end{table}

\subsection{SN quenching condition}

By substituting Eq.~\ref{eq:E+} and Eq.~\ref{eq:E-}, Eq.~\ref{eq:energyratecondition} can be cast in the following form:
\begin{subequations}
\begin{equation}
    {\rm SFR}\geq \frac{f_b M_h(z, T_{vir})}{\tau + \Delta t_{b}} \equiv {\rm SFR_{\rm min}},
    \label{eq:sfr_relation}
\end{equation}
where
\begin{equation}
   \tau = \frac{3}{2} \frac{\mu m_p G(\Delta t_{b}, Z_\star)}{n \Lambda(T_{vir}, Z)}\,.
\end{equation}
\end{subequations}
Eq.~\ref{eq:sfr_relation} describes the minimum star formation rate, $\rm SFR_{min}$, needed for SNe to suppress SF after a burst lasting for a time $\Delta t_{b}$. We note that the relation depends on the halo properties of the galaxy through $T_{vir}$ (or, equivalently, halo mass $M_h$). The metallicity also plays a key role, entering both in the gas cooling function and in the SNe energy input rate. Hereafter, for simplicity, we assume that $Z=Z_\star$.

When interpreting observations, by approximating the SFH of a galaxy before the quenching with a top-hat function based on SED fitting results, one can infer the corresponding values of $\rm SFR$ and $\Delta t_{b}$. If Eq.~\ref{eq:sfr_relation} is satisfied, SN feedback is predicted to halt SF in the galaxy; on the contrary, other mechanisms must be invoked.
Let us proceed with a more detailed analysis of this result by applying it to the two recently observed quiescent high-$z$ galaxies, whose properties are reported in Table~\ref{tab:gal_properties}.

\section{Results} \label{sec:results}
In order to compare observations with the SN-quenching condition illustrated by Eq.~\ref{eq:sfr_relation}, it is necessary to have information about the redshift, $z$, and the halo properties of the targeted galaxies, i.e., the halo mass, $M_h$, or the virial temperature, $T_{vir}(M_h,z)$ \cite[e.g.][]{BarkanaLoeb}.
Since it is difficult to directly determine $M_h$, particularly for high-$z$ galaxies, some assumptions need to be made to derive it. The most direct approach consists in considering the measured stellar mass of the galaxies and then derive the halo mass using a selected stellar mass - halo mass relation. For our analysis, we follow  \cite{Behroozi19}. The impact of this assumption is analyzed in Sec.~\ref{sec:sf_eff}.

\begin{figure*}[t!]
\centering
\includegraphics[width=\textwidth]{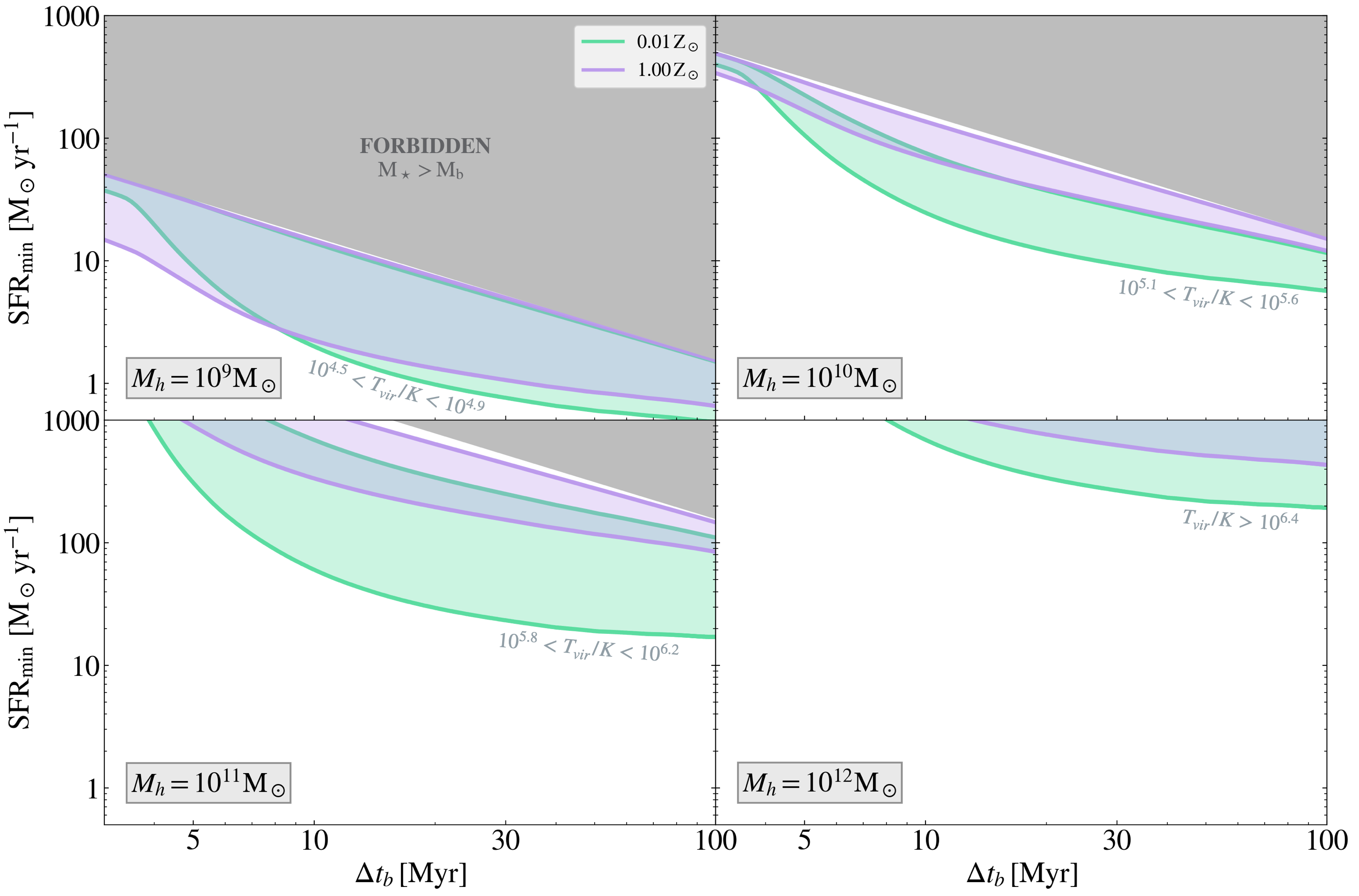}
\caption{
As Fig. \ref{fig:sfr_dt_jades} but for increasing halo mass from top-left ($M_h=10^9\msun$) to bottom-right ($M_h=10^{12}\msun$). The colors show two different gas metallicities, $Z=0.01 \zsun$ (green curves), $Z=\zsun$ (violet). The colored areas for each $Z$ value indicate redshift variations from $z=5$ (lower curve) to $z=15$ (upper); the corresponding  $T_{vir}$ is given in each panel.
\label{fig:sfr_dt_4Mh}
}
\end{figure*}

\subsection{Interpreting JADES-GS-z7-01-QU}

JADES-GS-z7-01-QU \citep{Looser23} has a redshift of $z=7.3$, and a stellar mass $M_\star = 10^{8.6} M_\odot$. Based on these measurements, using the stellar-to-halo mass relation by \cite{Behroozi19}, we find  $M_h = 10^{11} M_\odot$. This implies a virial temperature at that redshift of $T_{vir} = 10^{5.9}$ K (see Table~\ref{tab:gal_properties} for a summary of the galaxy's properties).
Having fixed these parameters, we plot ${\rm SFR_{min}}$ as a function of $\Delta t_b$ (Eq.~\ref{eq:sfr_relation}) for JADES-GS-z7-01-QU in the left panel of Fig.~\ref{fig:sfr_dt_jades}.

The curves shown in Fig.~\ref{fig:sfr_dt_jades} represent the value of $\rm SFR_{min}$ required for SNe to quench SF in the galaxy as a function of the burst duration, $\Delta t_{b}$. If a galaxy lies above the curve, after a timescale $\Delta t_{b}$ the SF will be quenched by SNe.
The gray region locates forbidden solutions for which $M_\star > f_b M_h(z, T_{vir})$. Also shown are curves for different metallicities\footnote{For the green $0.01\zsun$ curve, the metallicity value refers to the one assumed for the gas cooling rate. For the SNe energy input, since \code{starburst99} does not feature such low metallicity, we adopt the lowest available value of $0.05\,\zsun$.}. The increasing trend of $\rm SFR_{min}$ with metallicity indicates that low-$Z$ systems are more fragile to SN-induced quenching, due to their inability to efficiently cool the shocked gas and make it available again to form stars. As metallicity is increased from $Z=0.01 \zsun$ to solar value, $\rm SFR_{min}$ raises by $\simgt 1$ dex for a typical  $M_h=10^{11}\msun$.

Let us now directly compare the model with the SFH of JADES-GS-z7-01-QU. The post-starburst galaxy has experienced a short and intense burst of star formation followed by an abrupt SF drop. Its SFH, as inferred by \cite{Looser23} using \code{bagpipes} \citep{carnall_bagpipes}, can be approximated as a top-hat with $\rm SFR\sim 15~\msunyr$ and $\Delta t_{b} \sim 30~\rm Myr$. The measured stellar metallicity is $Z_\star \sim 10^{-2} \zsun$. Therefore, to guide the eye, the curve to compare the data with is the corresponding green one. Despite the low metallicity, we see that the JADES-GS-z7-01-QU point lies below the $\rm SFR_{min}$ curve by a factor $\sim 3$. This means that SN energy production {\it is not sufficient} to quench the star formation.
Therefore, we conclude that quiescence in JADES-GS-z7-01-QU must have been induced by a different  physical process. This finding lends further support to the previous studies based on cosmological simulations and quenching timescales reaching the same conclusion. In fact, not only the quenching appears to be too swift to be produced by SNe \citep{Gelli23, Dome23}, but our model convincingly shows that also the SFR level sustained in the observed burst cannot produce enough energy to suppress SF.

\subsection{Interpreting MACS0417-z5BBG}

MACS0417-z5BBG is a $z=5.2$ lensed galaxy observed by \cite{Strait23}; it has a low stellar mass, $M_\star =4.3\times 10^7\msun$. Just as JADES-GS-z7-01-QU, it exhibits a recent sudden SF drop, likely leading to a quiescent phase. 
Considering the SED fitting from \cite{Strait23}, the SFR is equivalent to a top hat with $\rm SFR\sim 1\,\msunyr$ and $\Delta t_{b}\sim40\rm~Myr$.
The estimated halo mass and virial temperature are $M_h=10^{10.5}\msun$ and $T_{vir} = 10^{5.5}$ K. Its stellar metallicity is rather high, $Z=\zsun$ (see Table~\ref{tab:gal_properties}). Thus in this case, to guide the eye, we should compare the data with the corresponding violet curve of Fig. \ref{fig:sfr_dt_jades}, right hand side panel, where the resulting ${\rm SFR_{min}}-\Delta t_{b}$ relation is shown.
We see that the galaxy falls by a factor $\sim 50$ below the critical $\rm SFR_{min}$ line, implying that SN energy cannot be responsible for the observed SF drop also in this case. Stated differently, SF could be quenched by SNe in this galaxy only if it would experience a $50\, \msunyr$ burst of SF. As for the previous case, a process different from SN explosions must be invoked to explain the data.

\subsection{General case} \label{sec:general}

To generalize our findings and enable their application to quiescent galaxies of different mass and redshift, we show the  ${\rm SFR_{min}}-\Delta t_{b}$ relation for a variety of cases in Fig.~\ref{fig:sfr_dt_4Mh}.
The panels correspond to four different halo masses in the range $M_h =10^{9-12}\msun$ , and two metallicities, $Z=0.01-1\, \zsun$. In order to also analyse the redshift dependence of SN quenching, the shaded areas in the figures cover, for each metallicity, the redshift range $5 < z < 15$, with $T_{\rm vir}=T_{\rm vir}(M_h,z)$ as indicated in each panel.

At low halo masses, SN quenching is more likely to happen as the conditions for star-formation suppression can be reached in short times and with low star formation rates.
For instance, for $M_h=10^9\msun$ (upper-left panel), the SNe produced in a burst of $\rm SFR>1\,\msun yr^{-1}$ are expected to quench the SF after only $\sim 20$~Myr at redshift $z\sim5$. In this case, there is a weak dependency on $Z$ descending from the small sensitivity of the cooling function on this quantity for $T_{vir} < 10^{4.9}$ K (see Fig.~\ref{fig:coolingrate}). The redshift dependence is instead more pronounced: for a $\sim10$~Myr burst, the $\rm SFR$ needed to quench SF increases from $\sim2 \, \msunyr$ at $z=5$ to $\sim20\, \msunyr$ at $z\sim15$.

At higher halo masses, meeting the quenching condition becomes more challenging (due to the increased gas mass that must be heated), and higher $\rm SFR$ values are required to suppress SF. The middle panels, displaying the $M_h=10^{10-11}\msun$ cases, illustrate the major role played by metallicity at intermediate virial temperatures, $T_{vir}\sim 10^{5} - 10^{6} $ K, with metal-poor galaxies being more likely to experience SN-quenching following bursts of star formation. 
The extreme case of $M_h=10^{12}\msun$ (bottom-right panel) essentially highlights the impossibility for very massive galaxies to be quenched by stellar feedback alone.

We can further test our model by applying it to observed bursty star-forming galaxies and verifying that they do not satisfy the quenching condition. Let us consider the case of GN-z11, the most UV-luminous galaxy at $z=10.6$ known to date \citep{Oesch16gn11, Bunker23gnz11}. It has a stellar mass of $M_\star = 5.4\times10^8\msun$, a halo mass of $M_h \sim 3\times 10^{10}~\msun$ \citep{Scholtz23gnz11} and gas metallicity of $\log(Z/\zsun)=-0.92$. Its SFH can be approximated with a $\rm SFR\approx20\msunyr$ burst lasting $\Delta t_b\approx 30$~Myr. These values are below the $\rm SFR_{min}$ threshold, confirming that SNe are not expected to quench the galaxy as expected.

\section{Discussion}\label{sec:discussion}

\subsection{Stellar-to-halo mass ratio dependence}\label{sec:sf_eff}
\begin{figure}[t]
\centering
\includegraphics[width=0.49\textwidth]{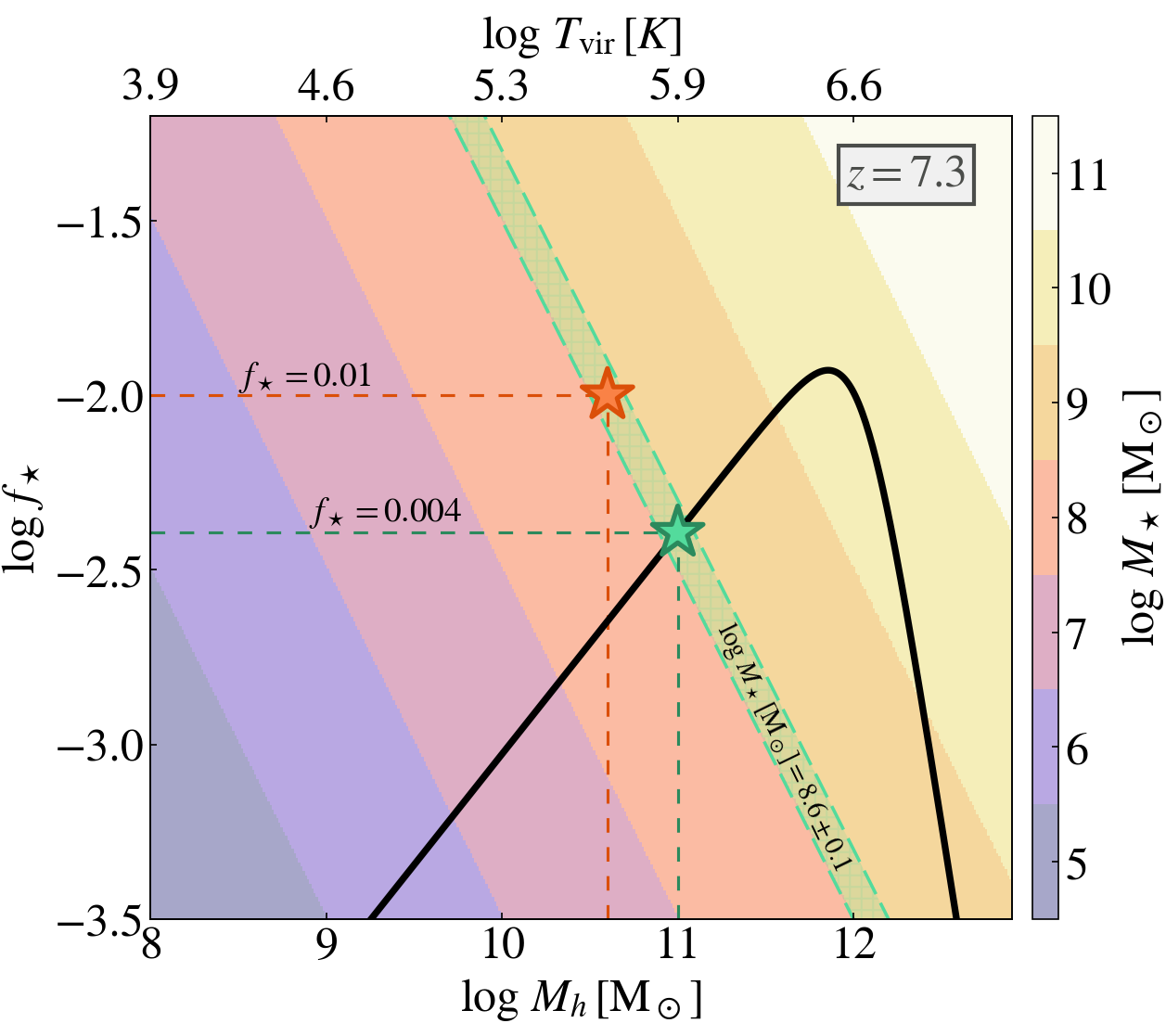}
\caption{
Stellar-to-halo mass ratio $f_\star= M_\star / M_h$ versus halo mass $M_h$, and corresponding virial temperature $T_{vir}$ (top axis), derived from \cite{Behroozi19} for redshift $z=7.3$ (black line). The colors illustrates the variation of the stellar mass $M_\star$; the green region identifies the measured stellar mass of JADES-GS-z7-01-QU, $M_\star = 10^{8.6}\msun$. Depending on the assumed ratio, different $M_h$ values can be obtained; for instance $f_\star = 0.004$ \citep[green star]{Behroozi19} yields $M_h=10^{11}\msun$, while for $f_\star=0.01$ $M_h=10^{10.6}\msun$ (orange star).
\label{fig:fstar_behroozi}
}
\end{figure}
\begin{figure}[t]
\centering
\includegraphics[width=0.48\textwidth]{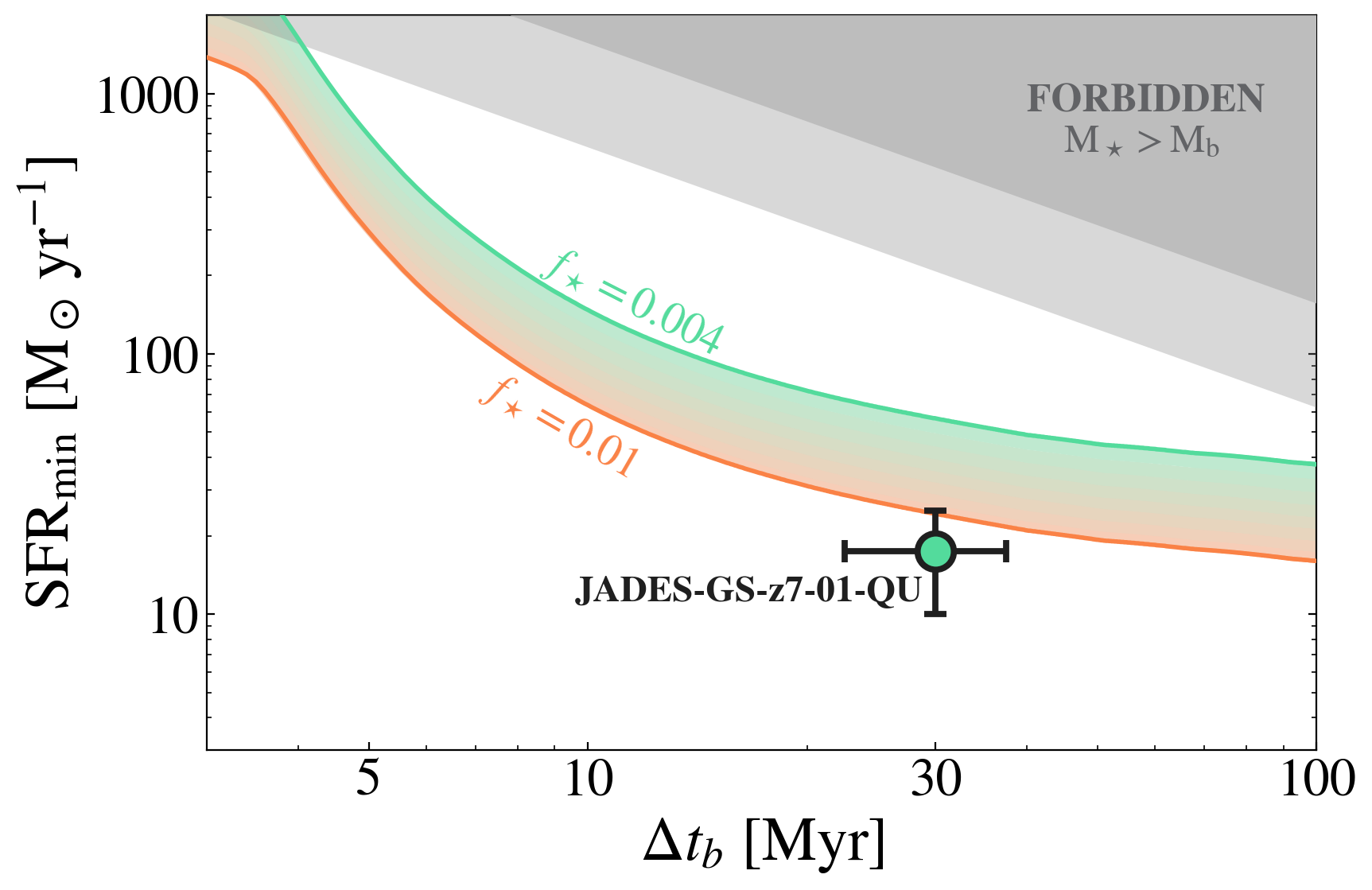}
\caption{
Variation of the ${\rm SFR}-\Delta t_{b}$ relation of Equation~\ref{eq:sfr_relation} at fixed redshift $z=7.3$ and metallicity $Z=0.01\zsun$ for different stellar-to-halo mass ratio.
The green curve corresponds to $f_\star=0.004$, the value derived for JADES-GS-z7-01-QU using \cite{Behroozi19} and already shown in Fig.~\ref{fig:sfr_dt_jades}. The orange curve is the one obtained when increasing the stellar-to-halo mass ratio to $f_\star=0.01$.
\label{fig:sfr_dt_jades_fstar}
}
\end{figure}

The SN-quenching condition illustrated by Eq.~\ref{eq:sfr_relation} depends on halo mass, metallicity and redshift. While the latter two quantities can be usually directly obtained from current observations, it is more difficult to measure $M_h$ at high-$z$. In the frequent case of unavailable kinematics measurements in the halo, some assumptions must be made in order to fix $M_h$.
This can be derived from the stellar mass by assuming a stellar-to-halo mass ratio defined as $f_\star = M_\star/M_h$. In the above analysis, we have adopted the redshift-dependent stellar-to-halo mass ratio vs halo mass relation derived by \cite{Behroozi19} through empirical modeling. However, this model is mostly based on constraints from galaxies observation at $z<8$ and $M_h>10^{11}\msun$. Indeed, especially at lower masses, the uncertain nature of the bursty SFR of high-$z$ galaxies makes it difficult to infer their stellar-to-halo mass relation from observations and, in general, simulations often predict higher SF efficiencies at high-$z$ \citep[e.g.][]{Xu16,Rosdahl2018,Pallottini2022}. We here want to assess the impact of the choice of $f_\star$.

To this aim, we refer to the case study of JADES-GS-z7-01-QU. In Fig.~\ref{fig:fstar_behroozi} we show the $f_\star - M_h$ relation by \cite{Behroozi19} at redshift $z=7.3$ (black line). The background colors show regions of constant stellar mass, with the green stripe locating the stellar mass of JADES-GS-z7-01-QU, $M_\star=10^{8.6}\msun$.
The green star marks the intersection with the \cite{Behroozi19} relation, thus identifying the values that we have been using for the observed galaxy: $f_\star=0.4\%$, $M_h=10^{11}\msun$, and $T_{vir}=10^{5.9}$ K.
Instead, assuming an higher value, e.g. $f_\star = 1\%$, the halo mass for JADES-GS-z7-01-QU is decreased to $M_h=10^{10.6}\msun$, or $T_{vir}=10^{5.7}$ K (orange star).

Fig.~\ref{fig:sfr_dt_jades_fstar} illustrates the impact on the ${\rm SFR_{min}}-\Delta t_{b}$ relation of these two different assumptions for $f_\star$. If $f_\star$ is increased, the consequently lower massive halo requires a smaller SFR to quench SF. The point corresponding to JADES-GS-z7-01-QU now lies closer to the curve, but it is still below the SN-quenching limit. This reinforces our findings meaning that, even when assuming a higher stellar-to-halo mass ratio, it is unlikely that the JADES-GS-z7-01-QU galaxy has been quenched solely by SNe. The action of SNe could be able to definitely provoke its quenching only if the ratio is as high as $f_\star\gtrsim 2\%$, and thus $M_h < 10^{10.4}\msun$.

Another factor of uncertainty is brought by the adopted stellar IMF. 
If the IMF at high-$z$ is more top-heavy than the assumed Salpeter, it would result in a larger  energy deposition by SNe \citep{Koutsouridou23}, hence facilitating galaxy quenching. Nonetheless, the metallicities in the two currently observed quiescent galaxies are sufficiently high ($Z>10^{-2}\zsun$) to rule out the presence of pristine stellar populations, typically associated with top-heavy IMFs.

\subsection{Interplay with other feedback processes}

From the analysis of the general trends of the SN-quenching condition, we have concluded that SF can be suppressed by SNe in: (a) low-mass ($M_h < 10^9\msun$) galaxies, or (b) highly star-forming galaxies, i.e., SFR $\gtrsim100~\msunyr$ for $M_h \sim 10^{11}\msun$ at $z\sim10$.

Specifically, for the two currently observed quenched high-$z$ galaxies JADES-GS-z7-01-QU and MACS0417-z5BBG, we find that the energy input from their SNe is not sufficient to produce a quenching. As a consequence, additional physical processes need to be at play in these systems on top of SNe to provide the extra energy needed to suppress star formation.

Environmental effects, such as ram pressure stripping and strangulation, can most likely be excluded as responsible for high-$z$ galaxies quenching since they are believed to occur predominantly at lower redshifts, once large-scale structures have developed in the Universe \citep[e.g.][]{Peng10}. Moreover, simulations predict that at high-$z$ the evolution and quenching of galaxies is driven by their mass rather than by the environment \citep{Contini20, Gelli20}. In the specific case of JADES-GS-z7-01-QU and MACS0417-z5BBG, furthermore, no massive nearby galaxy that would be able to contribute to environmental quenching is observed. Finally, these external physical processes act on typically long timescales \citep[$\gg 100\, \rm Myr$ e.g.][]{Emerick2016} and cannot cause a swift stop of star formation.

The additional energy needed to produce the witnessed quenching has to be instead associated with a fast physical mechanisms such as radiation-driven dusty outflows. 
The radiation pressure may be provided by young massive stars \citep[e.g.][]{Fiore23, Ziparo23, FerraraPallottiniDayal23} associated with the starburst. When the specific star formation rate exceeds a threshold, the galaxy becomes super-Eddington and develops powerful outflows clearing the galaxy and quenching the SF. The precise value of the sSFR threshold somewhat depends on the dust properties but a reasonable range is $\rm sSFR^{\star}=10-25\, \rm Gyr^{-1}$ \citep[Ferrara, in prep.]{Fiore23}. When such condition is satisfied, the outflows produced can efficiently contribute to heating and evacuating the gas available for star formation and lead to the galaxy quenching.
The sSFR derived for the starburst in JADES-GS-z7-01-QU and MACS0417-z5BBG are $43~\rm Gyr^{-1}$ and $25~\rm Gyr^{-1}$ respectively. Both galaxies exceed the sSFR threshold, implying that stellar radiation pressure may be a viable explanation for their quiescent state. This scenario is also supported by the low amount of dust ($A_V \sim 0.1$ and $\sim 0.2$ respectively), which should be  indeed efficiently ejected during the starburst phase \citep{FerraraPallottiniDayal23}.

Alternatively, the radiation pressure needed to quench the galaxy may be provided by an AGN. Even if supermassive black holes are typically invoked to explain suppression of star formation in massive galaxies ($M_\star > 10^{10}\msun$), there are some evidences of their presence also in $M_\star\approx 10^{8}\msun$ systems \citep[e.g.][]{Manzano-King19} and at high-$z$ \citep[e.g.][]{Maiolino23_gnz11}. However, in the cases of JADES-GS-z7-01-QU and MACS0417-z5BBG no broad emission lines are observed, implying that the accretion stops as soon as the star formation does, and the putative black hole must be dormant in these galaxies at the time of observation.

\section{Conclusions}\label{sec:conclusions}

The recent JWST discoveries of two high-$z$ post starburst quiescent systems prompted us to investigate the role of SN-feedback in driving galaxies quenching.
Through a minimal physical model we identify the conditions under which the SNe explosions associated with a burst of star formation can lead to a total quenching, bringing the galaxy to a temporary quiescent state. We found that:
\begin{itemize}
    \item The minimum $\rm SFR$ required for SNe to quench SF,  ${\rm SFR_{min}}$, can be expressed as a function of the burst duration, $\Delta t_{b}$;
    \item Such ${\rm SFR_{min}}-\Delta t_{b}$ relation depends on the halo mass $M_h$, metallicity $Z$ and redshift $z$; lower mass, more metal-poor, lower redshift systems are more easily quenched by stellar feedback;
    \item  SNe cannot be responsible for the quenching of JADES-GS-z7-01-QU \cite{Looser23} and MACS0417-z5BBG \cite{Strait23} since the energy produced in the burst is not sufficient to suppress star formation;
    \item  Given the high specific SFR of the two quiescent galaxies, corresponding to super-Eddington luminosities,  we suggest that quenching may instead be produced by radiation-driven dust outflows. 
\end{itemize}

We predict that with the incoming JWST data, an increasing number of quiescent galaxies will be discovered at high redshifts, particularly among low-mass galaxies, due to their bursty nature. The derived $\rm SFR_{min}-\Delta t_{b}$ relation may serve as a valuable tool for interpreting the SFHs of observed post-starburst quenched galaxies, allowing us to solidly determine the role of SN feedback in rapid SF suppression.

\begin{acknowledgments}
This project received funding from the ERC Starting Grant NEFERTITI H2020/804240 (PI: Salvadori). AF acknowledges the ERC Advanced Grant INTERSTELLAR H2020/740120 (PI: Ferrara). We acknowledge the computational resources of the Center for High Performance Computing (CHPC) at SNS.
\end{acknowledgments}

\bibliography{refer,codes}
\bibliographystyle{aasjournal}

\end{document}